\documentclass[aps,reprint,prl,showpacs,amsmath,floatfix,superscriptaddress]{revtex4-1}

\usepackage{graphicx}
\usepackage{bbold}
\usepackage{times}
\usepackage{hyperref}
\usepackage{dsfont}
\usepackage{color}
\begin{document}

\title{Steady States of Infinite-Size Dissipative Quantum Chains via Imaginary Time Evolution}

\author{Adil A. Gangat}
\affiliation{Department of Physics, National Taiwan University, Taipei 10607, Taiwan}
\author{Te I}
\affiliation{Department of Physics, National Taiwan University, Taipei 10607, Taiwan}
\author{Ying-Jer Kao}
\email{yjkao@phys.ntu.edu.tw}
\affiliation{Department of Physics, National Taiwan University, Taipei 10607, Taiwan}
\affiliation{National Center of Theoretical Sciences, National Tsinghua University, Hsinchu 30013, Taiwan}
\date{\today}
\begin{abstract}
Directly in the thermodynamic limit, we show how to combine imaginary and real time evolution of tensor networks to efficiently and accurately find the nonequilibrium steady states (NESS) of one-dimensional dissipative quantum lattices governed by the Lindblad master equation.
The imaginary time evolution first bypasses any highly correlated portions of the real-time evolution trajectory by directly converging to the weakly correlated subspace of the NESS, after which real time evolution completes the convergence to the NESS with high accuracy.  
We demonstrate the power of the method with the dissipative transverse field quantum Ising chain. We show that a crossover of an order parameter shown to be smooth in previous finite-size studies remains smooth in the thermodynamic limit.

\end{abstract}

\maketitle

\textit{Introduction}--The out-of-equilibrium behavior of dissipative many-body systems is of relevance to experimental platforms such as trapped ions\cite{schindler2013quantum}, cold atoms \cite{PhysRevA.87.022110}, superconducting circuits\cite{Houck2012-cu,schmidt2013circuit,hur2015many}, and nanoelectromechanical systems \cite{lozada2016quantum}. 
Theoretical activity has recently increased around developing numerical methods for determining the non-equilibrium steady states (NESS) of such dissipative lattices \cite{Werner:2016gi,Weimer2015-is, Cui2015-ht,Finazzi2015-ml,Mascarenhas2015-vw,Dorda2015-zt,Raghu:2016sh,jin2016cluster,Roman_new}. 
 This includes tensor network methods \cite{Schollwock2011-hd,Verstraete2008-zy,Cirac2009-dg,Orus2014-ee}, which serve as an efficient numerical ansatz for states that obey an area law\cite{Eisert2010-ea} (i.e. have small correlations between their bipartitions). 
It is proved that the mutual information of the NESS of a local dissipative quantum system satisfies an area law~\cite{Brandao2015-pq,mahajan2016entanglement}, assuring that their tensor network representation, if found, will be computationally efficient.
 Also, a proof for the stability\cite{Cubitt2015-gi} of NESS against local system perturbations assures that theoretically determined NESS of translationally invariant systems are of relevance to experiments, where translational invariance can only be approximate.  

The focus of the present work is on the problem of numerically finding the tensor network representation of the NESS of translationally invariant one-dimensional systems in the thermodynamic limit.  
The infinite-size version of the time-evolving block decimation (iTEBD) algorithm \cite{Vidal2007-px} enables a (local) time-evolution directly in the thermodynamic limit using a matrix product state (MPS) that spans only a single unit cell.  
Also, the iTEBD algorithm can be used for both imaginary and real time evolution with local operators.

When imaginary time evolution is performed with a Hamiltonian, the fixed point lies in the ground state manifold of the given Hamiltonian. 
An efficient tensor network can accommodate the entire imaginary time evolution trajectory if the fixed point obeys an area law. 
Furthermore, the convergence toward the fixed point is exponentially fast with a rate proportional to the energy gap of the Hamiltonian. 
Therefore, imaginary time evolution with iTEBD is a very efficient way to obtain  ground states of local Hamiltonians in the thermodynamic limit.  

For real time evolution of dissipative systems, the fixed point lies in the NESS manifold. 
However, real time evolution with iTEBD~\cite{Orus2008-pz} is not always an efficient way to obtain the NESS since portions of the real time evolution trajectory from the initial state to the NESS may not obey an area law even when the NESS does so itself~\cite{Cui2015-ht}. 
Further, the rate of convergence to the NESS may be very slow~\cite{Mascarenhas2015-vw,PhysRevLett.111.150403}.  
For finite-size chains, recent works\cite{Cui2015-ht,Mascarenhas2015-vw} tackle these problems using variational methods. 
These approaches, though accurate, can not be easily generalized to infinite-size systems.

All prior approaches to finding groundstates and NESS with tensor networks have exploited the area law property of the target state to achieve efficiency. By additionally exploiting a closely related but distinct physical property, exponential decay of two-point correlators, we here show how to alleviate the inefficiencies of real time evolution to the NESS for \textit{infinite-size} one-dimensional tensor networks: we construct an auxiliary local Hamiltonian such that its ground state approximates the NESS, and perform iTEBD imaginary time evolution with the auxiliary Hamiltonian. 
In this way, we are able to bypass any highly correlated portions of the real time evolution trajectory and to arrive in the area law-obeying neighborhood of the NESS exponentially quickly.
We further improve the convergence to the NESS using a real time evolution of the Lindbladian.

The auxiliary Hamiltonian approach that we present holds potential for the thermodynamic limit of higher dimensional dissipative systems as well, since both imaginary time evolution and variational optimization \cite{PhysRevB.94.035133} have been demonstrated to obtain ground states with the higher dimensional tensor network ansatz of infinite-size projected entangled pair states (iPEPS) \cite{Jordan2008-sc}.

As a demonstration of our method, we use it to probe the thermodynamic limit of a crossover in an order parameter of the one-dimensional dissipative transverse field Ising model. 
To our knowledge, non-mean-field studies of this crossover have only previously been performed in the finite-size limit. 
Our result shows that this crossover remains smooth in the thermodynamic limit.

\textit{Method}--The equation of motion for a (discrete) quantum system coupled to a Markovian environment is given by the Lindblad master equation (LME) \cite{Breuer2002-vc}.  Under the Choi isomorphism ($\hat{\rho}=\sum_j p_j|\Psi_j\rangle\langle\Psi_j| \rightarrow |\rho\rangle=\sum_j p_j |\Psi_j\rangle\otimes|\Psi_j\rangle$), the LME is
\begin{equation}
\frac{d}{dt}|\rho\rangle=\hat{\mathcal{L}}|\rho\rangle,
\label{rME}
\end{equation}
with
\begin{align}
\hat{\mathcal{L}}=&-\frac{i}{\hbar}(H\otimes\mathbb{1}-\mathbb{1}\otimes H^{\textrm{T}}) \nonumber\\&+\sum_{\alpha}\frac{1}{2}(2L_{\alpha}\otimes\bar{L}_{\alpha}-L_{\alpha}^{\dagger}L_{\alpha}\otimes\mathbb{1}-\mathbb{1}\otimes L_{\alpha}^{\textrm{T}}\bar{L}_{\alpha}),
\label{Lindbladian}
\end{align}
where $H$ is the system Hamiltonian and $L_{\alpha}$ are dissipative operators.  

A matrix product density operator (MPDO)\cite{Verstraete2004-hc} is a tensor network that serves as an efficient ansatz for a mixed state density matrix of a one-dimensional lattice when the correlations between real space bipartitions of the state are small \cite{Zwolak2004-tz, Verstraete2004-hc}. 
Under the Choi isomorphism, the MPDO \cite{Verstraete2004-hc} can be written as a matrix product state (MPS) \cite{Perez-Garcia2007-ep}: $|\rho\rangle=\sum_{s_1,...,s_N=1}^d \textrm{Tr}(A_1^{s_1}...A_N^{s_N})|s_1...s_N\rangle$, where the $A_j$ are tensors of dimension $d \times D \times D$ and $D$ is referred to as the ``bond dimension". 
The chief advantage of such an ansatz is that while $D$ needs to be exponentially large in the system size for the MPS to be exact, a much smaller (i.e. computationally tractable) value of $D$ yields extremely high accuracy for states that obey an area law \cite{Eisert2010-ea}. 
 Another advantage of this ansatz is that the entanglement spectrum (denoted $\lambda_i^2$ below) between subblocks of any bipartition of the lattice is readily calculated \cite{Vidal2004-ru}.  
In the case of infinite-size systems with translational invariance, the MPS or MPDO is referred to as an ``iMPS" or ``iMPDO", and it can span a Hilbert space as small as a single unit cell of the target physical state \cite{Vidal2007-px}.

The NESS (denoted $|\rho_{\infty}\rangle$) of dissipative systems described by the LME are defined by $\hat{\mathcal{L}}|\rho_{\infty}\rangle=0$, with the constraint that the $|\rho_{\infty}\rangle$ are vectorized forms of positive operators with unit trace (i.e. physically valid density matrices; see below for further discussion).  
The authors of Ref.~\cite{Cui2015-ht} observe that $|\rho_{\infty}\rangle$ will also be the ground state of the \textit{nonlocal} Hamiltonian $\hat{\mathcal{L}}^{\dagger}\hat{\mathcal{L}}$. 
For finite-size chains, they present a variational method for finding the (non-degenerate) ground state of $\hat{\mathcal{L}}^{\dagger}\hat{\mathcal{L}}$  as a way of determining $|\rho_{\infty}\rangle$.  
However, their method does not apply directly (i.e. without costly extrapolation from finite-size scaling) in the thermodynamic limit. Since $\hat{\mathcal{L}}^{\dagger}\hat{\mathcal{L}}$ is nonlocal, imaginary time evolution with iTEBD also can not be used to find its ground state. 
Here we show that it is possible to construct a \textit{local} auxiliary Hamiltonian $\mathcal{H}$ such that the iTEBD algorithm may be used to approach the infinite-size NESS via imaginary time evolution:
\begin{align}
|\rho_{\infty}\rangle \approx \lim_{\tau\rightarrow\infty} \frac{\textrm{exp}(-\mathcal{H}\tau)|\rho_0\rangle}{||\textrm{exp}(-\mathcal{H}\tau)|\rho_0\rangle||},
\label{itime}
\end{align}
where $|\rho_0\rangle$ is any vectorized density matrix such that $\langle\rho_{\infty}|\rho_0\rangle\neq0$. 

As an initial motivation we observe that if $F$ is a \textit{local} Hamiltonian with positive eigenvalues, $F^2$ will be a \textit{nonlocal} Hamiltonian with the same ground and excited states as the local $F$. 
This suggests the possibility of finding a local Hamiltonian $\mathcal{H}$ whose ground state is at least a good approximation to the ground state of the nonlocal $\hat{\mathcal{L}}^{\dagger}\hat{\mathcal{L}}$.

We assume that $\hat{\mathcal{L}}$ is a translationally invariant local operator; this corresponds to the case of a translationally invariant local system Hamiltonian $H$ and translationally invariant local dissipation operators $L_{\alpha}$.  $\hat{\mathcal{L}}$ can therefore be expressed as a sum of translationally invariant local terms ($\hat{\mathcal{L}}=\sum_{r\epsilon\mathbb{Z}} \hat{\mathcal{L}}_r$) and we may write
\begin{align}
\hat{\mathcal{L}}^{\dagger}\hat{\mathcal{L}}=\sum_{r,s\epsilon\mathbb{Z}}\hat{\mathcal{L}}_{r}^{\dagger}\hat{\mathcal{L}}_{s}.
\label{fullexpansion}
\end{align}
We may use $\langle\hat{\mathcal{L}}_r\rangle=0$ as a benchmark, and $|\langle\hat{\mathcal{L}}_r\rangle|$ as a figure of merit.  
It has been analytically proven in Ref.~\cite{Brandao2015-pq} that the two-point correlator shows an exponential decay in NESS.
Therefore the long-range couplings in $\hat{\mathcal{L}}^{\dagger}\hat{\mathcal{L}}$ do not play a significant role in determining its ground state $|\rho_{\infty}\rangle$ and we may truncate the second sum in Eq.~(\ref{fullexpansion}) by setting $s=r$.  
Further taking the $k^{th}$ root of each remaining term we arrive at the proposed local auxiliary Hamiltonian for imaginary time evolution to the NESS:
\begin{align}
\mathcal{H}=\sum_{r\epsilon\mathbb{Z}}\big(\hat{\mathcal{L}}_{r}^{\dagger}\hat{\mathcal{L}}_{r}\big)^{1/k}.
\label{auxHam}
\end{align}
If the (numerical) gap between the lowest two eigenvalues of $\hat{\mathcal{L}}_{r}^{\dagger}\hat{\mathcal{L}}_{r}$ is less than one, $k>1$ will increase the gap since $\hat{\mathcal{L}}_{r}^{\dagger}\hat{\mathcal{L}}_{r}$ is positive semi-definite. 
This will yield faster convergence toward the ground state of $\mathcal{H}$ (for imaginary time evolution the rate of convergence is proportional to the gap).  
While $\mathcal{H}$ for $k>1$ will not generally commute with $\mathcal{H}$ for $k=1$, we find that the advantage gained (see next section) outweighs this drawback.

The form of $\hat{\mathcal{L}}_r$ can change as long as $\hat{\mathcal{L}}=\sum_{r\epsilon\mathbb{Z}} \hat{\mathcal{L}}_r$. $\hat{\mathcal{L}}_r$ with larger support leads to longer range couplings in $\mathcal{H}$; for $\hat{\mathcal{L}}_r$ of infinite support, $\mathcal{H}$ becomes equal to $\hat{\mathcal{L}}^{\dagger}\hat{\mathcal{L}}$ (when $k=1$).  
We may therefore decrease the distance between $|\rho_{\infty}\rangle$ and the ground state of $\mathcal{H}$ by increasing the support of $\hat{\mathcal{L}}_r$, albeit with an increased computational cost.

Rather than relying on imaginary time evolution alone, the following hybrid method can be more efficient: with small $D$, relatively large timestep, and $\hat{\mathcal{L}}_r$ of small support, imaginary time evolution can be used to rapidly converge to the area law-obeying neighborhood of the NESS, after which real time evolution with iTEBD can minimize $|\langle\hat{\mathcal{L}}_r\rangle|$ with successively smaller timesteps and successively larger $D$.  
The Trotter error of iTEBD vanishes exponentially in the timestep size, and the error associated with a finite $D$ vanishes exponentially in $D$ when there is an area law.  

As mentioned, for the converged solution to be physically valid it must correspond to a positive operator. 
If the NESS is unique, it is shown in Ref.~\cite{Cui2015-ht} that this requirement is 
automatically satisfied when $\hat{\mathcal{L}}|\rho\rangle=0$ . 
We may therefore assume a unique NESS and assure (near-)positivity by attaining very small $|\langle\hat{\mathcal{L}}_r\rangle|$ and converging the spectrum. 
This is similar to what is done in Ref.~\cite{Cui2015-ht}, which assures (near-)positivity by variationally reducing $\langle\hat{\mathcal{L}}^{\dagger}\hat{\mathcal{L}}\rangle$ to below a chosen threshold and converging various observables, and Ref.~\cite{Mascarenhas2015-vw}, which assures (near-)positivity by variationally minimizing $||\hat{\mathcal{L}}|\rho\rangle||$.
As an alternative, which we do not  implement here, positivity may be enforced even in the case of non-unique NESS by projecting $|\rho\rangle$ onto the physical subspace. We outline this procedure in the Appendix.  

\textit{Numerical Results--}There have been various numerical investigations of the 1D quantum Ising model with nearest-neighbor coupling, uniform transverse magnetic field, and on-site dissipation \cite{Cui2015-ht, Werner2005-ct, werner2005quantum, yin2014nonequilibrium, orth2008dissipative, PhysRevA.87.022110, PhysRevLett.111.150403, Hu2013-yo,Ates2012-hj,Lee2011-xj,Joshi2013-fb,marcuzzi2014universal, Weimer2015-wq, Rose2016-fw}. 
Here we look at the following incarnation: the system is governed by Eqs. (\ref{rME}) and (\ref{Lindbladian}) with the Hamiltonian given by $H=\sum_{j\epsilon\mathbb{Z}} H_j$, where ($\hbar=1$)
\begin{align}
H_j=\sigma_z^{[j]}\sigma_z^{[j+1]}+h_x\sigma_x^{[j]},
\label{Ising}
\end{align}
$j$ is the lattice site index, and $h_x$ is the transverse field strength; the local dissipation terms are given by $L_{j}=\sqrt{\gamma}\sigma_-^{[j]}$, where $\gamma$ is a decay rate from spin up to spin down.  

We choose a tensor network that is structured such that two physical sites are associated to each of the numerical sites of a two-site iMPDO.  
For both imaginary and real time evolution, a fourth order Suzuki-Trotter expansion of the evolution operators is used.  
For the imaginary time evolution, we choose the following 4-local form ($\hbar=1$) for $\hat{\mathcal{L}}_r$:
\begin{align}
&\hat{\mathcal{L}}_{r} = -i(H_r\otimes\mathbb{1}-\mathbb{1}\otimes H_r^{\textrm{T}}) \nonumber\\&~~~~~~~+ \frac{1}{2}\textrm{diss}[2r] + \frac{1}{2}\textrm{diss}[2r+1] \nonumber\\ &~~~~~~~+ \frac{1}{2}\textrm{diss}[2r+2] + \frac{1}{2}\textrm{diss}[2r+3],\nonumber\\
&H_r=\frac{1}{2}\sigma_z^{[2r]}\sigma_z^{[2r+1]}+\sigma_z^{[2r+1]}\sigma_z^{[2r+2]}+\frac{1}{2}\sigma_z^{[2r+2]}\sigma_z^{[2r+3]} \nonumber\\&~~~~~~~+\frac{1}{2}h_x(\sigma_x^{[2r]}+\sigma_x^{[2r+1]}+\sigma_x^{[2r+2]}+\sigma_x^{[2r+3]}),
\label{4local}
\end{align}
where the local dissipation term is given as $\textrm{diss}[r]=\frac{\gamma}{2}(2\sigma_{-}^{[r]}\otimes\sigma_{-}^{[r]}  - \sigma_{+}^{[r]}\sigma_{-}^{[r]}\otimes\mathbb{1}^{[r]}-\mathbb{1}^{[r]}\otimes \sigma_{+}^{[r]}\sigma_{-}^{[r]})$. 
The overlap between $\hat{\mathcal{L}}_r$ and $\hat{\mathcal{L}}_{r+1}$ in this form is two sites; forms resulting in three site overlap would yield more intersite couplings in $\mathcal{H}$ and therefore a better match between the ground state of $\mathcal{H}$ and $|\rho_{\infty}\rangle$, but the above form is numerically more efficient and sufficient for our demonstration.  
It is straightforward to verify that this form satisfies $\hat{\mathcal{L}}=\sum_{r\epsilon\mathbb{Z}} \hat{\mathcal{L}}_r$. 
 
\begin{figure}
\includegraphics[scale=0.37]{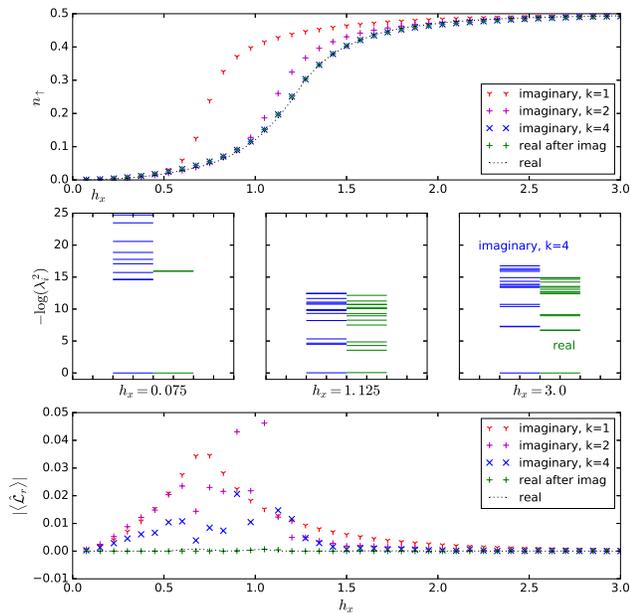}
\caption{(color online).  A comparison of iTEBD imaginary time and iTEBD real time evolution, and a demonstration of the hybrid method, for an infinite-size Ising chain with Hamiltonian of the form in Eq. (\ref{Ising}) and dissipation rate $\gamma=0.5$ when the spectrum is converged as described in the main text.  The results show that imaginary time evolution can serve as an alternative to real time evolution for approaching the NESS.  The imaginary time evolution results are for the 4-local form of $\hat{\mathcal{L}}_r$ given in Eq.(\ref{4local}).  (Top) Converged up-spin density $n_{\uparrow}$ vs. transverse magnetic field strength $h_x$ for real time evolution and different values of $k$ for the imaginary time evolution. The overall accuracy of the imaginary time evolution improves with larger $k$.  The data for the hybrid method indicates that performing real time evolution after the imaginary time evolution enhances the accuracy as expected. (Middle) Converged entanglement spectra for three different values of $h_x$, with imaginary time evolution results shown for $k=4$. The rapid decay of the spectra shows that both evolutions converge to a weakly correlated subspace, suggesting proximity to the NESS. (Bottom) For real time and imaginary time with $k=4$, $|\langle \hat{\mathcal{L}}_r\rangle|$ converges to values much less than the smallest energy scale in the system, also suggesting proximity to the NESS.}
\label{fig_4localupspin}
\end{figure}

For the finite-size version of this model, a previous numerical study~\cite{Hu2013-yo} shows that a smooth crossover occurs in the value of the up-spin density $n_{\uparrow}=\sum_r\langle \hat{n}_{\uparrow}^{[r]} \rangle/N$ as $h_x/\gamma$ is increased from small to large values.  
 To demonstrate the theoretical validity of the imaginary time method put forth in the previous section, we probe this crossover (for $\gamma=0.5$) in the thermodynamic limit with imaginary time evolution and real time evolution independently.  
 The purpose here is to demonstrate that the imaginary time method can come close to the NESS very efficiently and robustly (i.e. with very crude parameters and convergence).  
In the Appendix, we check the convergence of the imaginary time evolution vs. both $D$ and imaginary time step size at the middle of the crossover.  
From there we determine that fixing $D=15$ and the imaginary timestep size at $d\tau=10^{-2}$ (in units of the inverse Ising interaction strength squared) is sufficient for this purpose. 
For real time evolution we also use a timestep size of $dt=10^{-2}$ (in units of inverse Ising interaction strength).  
The simulations are run until the spectrum is converged according to the following criterion: $|\lambda_{1}(t+\Delta t) - \lambda_{1}(t)|/\lambda_{1}(t+\Delta t) < 10^{-7}$, where $\lambda_{1}$ is the largest singular value in the iMPDO.  
At each value of $h_x$ the same initial state is used for both imaginary time and real time evolution.
The results are illustrated in Fig. \ref{fig_4localupspin}.  
The spectra in the middle panel shows that the imaginary time evolution ($k=4$) and the real time evolution have converged to a weakly correlated subspace.  
The bottom panel shows that $|\langle\hat{\mathcal{L}}_r\rangle|$ is much smaller than the smallest energy scale in $\hat{\mathcal{L}}_r$ for real time evolution and imaginary time evolution (when $k=4$).
The results in these two panels indicate proximity to the NESS.  
Because $|\langle\hat{\mathcal{L}}_r\rangle|$ is smallest for real time evolution, the data in the top panel for the real time evolution can be considered the most accurate, and the increasing overall match between the imaginary time data and real time data for $n_{\uparrow}$ with higher $k$ shows that the accuracy of the imaginary time evolution improves with larger $k$.  
Though we do not investigate it here, it is possible that the significant enhancement in accuracy with larger $k$ is due to the removal of metastability \cite{Rose, Macieszczak2016-wq}. 
Taken together, these results demonstrate the conceptual validity of the imaginary time evolution method as an alternative to real time evolution for approaching the state space neighborhood of the NESS.
The level of accuracy achieved with the imaginary time method is remarkable given the crude convergence scheme and enormous truncation of $\hat{\mathcal{L}}^{\dagger}\hat{\mathcal{L}}$ in forming $\mathcal{H}$.  
The physical explanation for the success of the truncation of $\hat{\mathcal{L}}^{\dagger}\hat{\mathcal{L}}$ is the exponential decay of the two point correlators in the NESS.  
For comparison, we show simulation results for a 2-local form of $\hat{\mathcal{L}}_r$ in the Appendix.  The 2-local form also permits successful imaginary time evolution toward the NESS, but with poorer accuracy.

We note that our results indicate that in this one-dimensional case the crossover in $n_{\uparrow}$ remains smooth in the thermodynamic limit; this is in contrast to the sharp transition predicted for the two-dimensional version of the model \cite{Weimer2015-is, Weimer2015-wq}.
This is the first study of this crossover  in the thermodynamic limit beyond the mean-field theory.

We also demonstrate the functionality of the full hybrid method (imaginary time evolution followed by real time evolution).  
For the same system parameters as the other data, the top panel of Fig.~\ref{fig_4localupspin} shows the converged data from the hybrid method when the state at the end of the imaginary time evolution for $k=4$ is converged to $|\langle \hat{\mathcal{L}}_r \rangle|<10^{-3}$ with real time evolution.

To demonstrate that the hybrid method remains efficient where real time evolution becomes inefficient, with $\gamma=0.01$ and $h_x=1$, we evolve a highly correlated initial state with the two methods separately and find that the hybrid method converges rapidly in several minutes of computation time with $D=6$ while real time evolution over the same number of timesteps becomes trapped in an area law-violating sector of the Hilbert space even with $D$ as high as $25$.  
The time evolution of $|\langle \hat{\mathcal{L}}_r \rangle|$ is shown in Fig.~\ref{fig_hybridmain}, while the evolution of the spectra and further details are given in the Appendix.
It is clear that the initial imaginary time evolution bypasses the highly correlated region of the real-time trajectory, and brings the iMPDO close to the neighborhood of the NESS; thus, an efficient real time evolution becomes possible, and very high accuracy can be achieved. 

\begin{figure}
\includegraphics[scale=0.38]{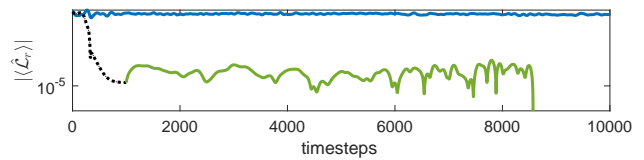}
\caption{(color online). System parameters: $\gamma=0.01$, $h_x=1$.  Real time evolution (upper solid line, blue) with $D$=25 from the initial state fails to converge to the NESS over a timescale equal to the largest intrinsic timescale of the system.  With the hybrid method method  ($D$=6), imaginary time evolution (dotted line) rapidly achieves proximity to the NESS, after which real time evolution (lower solid line, green) efficiently attains $|\langle \hat{\mathcal{L}}_r \rangle|<10^{-6}$.}
\label{fig_hybridmain}
\end{figure}

\textit{Discussion--} Imaginary time evolution of tensor networks with iTEBD is an efficient way of approximating the ground states of many-body quantum systems in the thermodynamic limit when an area law holds for the ground state; the area law permits high accuracy with a computationally tractable bond dimension. 
Here we have shown that imaginary time evolution of an iMPDO with iTEBD is an efficient way of approximating the NESS of dissipative quantum chains when the NESS has both an area law and exponential decay of two-point correlators; the area law permits the iMPDO to approximate the NESS with high accuracy with a computationally tractable bond dimension, and the exponential decay of two-point correlators permits the construction of a \textit{local} auxiliary Hamiltonian with which to perform imaginary time evolution with iTEBD. 
We have demonstrated that imaginary time evolution can very efficiently reach the area law-obeying state space neighborhood of the NESS. 
Real time evolution can be employed after the imaginary time evolution to efficiently obtain the NESS with very high accuracy. 
 Alternatively, but perhaps less efficiently, imaginary time evolution may be used alone to reach the NESS by converging not only in the timestep size and bond dimension, but also the support size of the local terms in the auxiliary Hamiltonian.

In the course of demonstrating our method with the dissipative transverse field quantum Ising chain, we have shown that a crossover of an order parameter that is smooth in the finite-size limit remains smooth in the thermodynamic limit, in contrast to the two-dimensional case.

With the higher dimensional tensor network of iPEPS, both variational optimization and imaginary time evolution have been shown to work \cite{Jordan2008-sc,PhysRevB.94.035133}, and real time evolution toward NESS was also recently demonstrated \cite{Roman_new}; the method presented here may thereby be extended to higher dimensions.

We note that the imaginary time evolution method presented here can also apply to finite-size chains by using TEBD instead of iTEBD, and it would be interesting to compare the performance of the hybrid method in this paper with the variational methods\cite{Cui2015-ht,Mascarenhas2015-vw} for finding the NESS of finite-size chains.

%Finally, we note that the Lindbladian superoperator of Eq.~(\ref{Lindbladian}) can be written in terms of matrix product operators; therefore it is possible to construct a full auxiliary Hamiltonian $\mathcal{H}=\hat{\mathcal{L}}^{\dagger}\hat{\mathcal{L}}$ and use an infinite DMRG algorithm\cite{McCulloch:2008fv} to do direct variational optimization of the iMPDO\cite{Te-I:oq}.

YJK and AAG acknowledge discussions with Ian McCulloch, Roman Or\'{u}s and Augustine Kshetrimayum, and the hospitality of the Institut f\"{u}r Physik at Johannes Gutenberg-Universit\"{a}t.  AAG acknowledges early pedagogical interactions at the University of Queensland with Guifr\'{e} Vidal and Ho N. Phien regarding MPS and TEBD. We thank Dominic C. Rose for insightful remarks regarding metastability.  Some of the simulations were coded using the Uni10 tensor library~\cite{Kao:2015gb}. This work is supported by the MOST in Taiwan through Grants No.~104-2112-M-002 -022 -MY3, 105-2112-M-002 -023 -MY3.

\bibliographystyle{apsrev4-1}
\bibliography{Ref}

\section{APPENDIX}

\subsection{Positivity enforcement via projection onto physical subspace}

We outline here a more rigorous method to ensure positivity of the state vector $|\rho\rangle$.  A basis of physical state vectors can be constructed by performing the Choi isomorphism on a set of random physical density matrices (which may or may not be orthonormalized): $\{\tilde{\phi_i}=|\phi_i\rangle\langle \phi_i|\}\rightarrow\{|\tilde{\phi_i}\rangle=|\phi_i\rangle\otimes|\phi_i\rangle\}$. (The constructed physical basis may be optimized by restricting to basis states that obey an area law.) Then $|\tilde{\rho}\rangle=(\sum_i |\tilde{\phi_i}\rangle\langle \tilde{\phi_i}|)|\rho\rangle$ will correspond to a positive operator.  Normalization, if needed, can be performed as mentioned in Ref. [\cite{Cui2015-ht}].  Such a projection may, however, increase $|\langle \hat{\mathcal{L}}_r \rangle|$, especially since the constructed physical basis will not necessarily span the entire physical subspace. If the increase in $|\langle \hat{\mathcal{L}}_r \rangle|$ after the projection is too large, the minimization of $|\langle\hat{\mathcal{L}}_r\rangle|$ may be resumed with $|\tilde{\rho}\rangle$, or different random physical bases may be tried until one is found that gives a satisfactory result for $|\langle\tilde{\rho}|\hat{\mathcal{L}}_r|\tilde{\rho}\rangle|$.

\subsection{Convergence test of bond dimension and timestep size for imaginary time evolution}

Here we show results for convergence tests of the imaginary time evolution method with the 4-local form of $\hat{\mathcal{L}}_r$ (see main text). For $\gamma=0.5$, the middle of the crossover in $\langle n_{\uparrow} \rangle$ occurs at about $h_x=1.2$, and we perform the tests with these parameters fixed at these values.  

The first test assesses convergence vs. the bond dimension $D$ with a fixed timestep size of $\tau=10^{-2}$.  A random initial state is chosen as the initial condition for the simulation with the smallest value of $D$.  When the simulation is stopped, $D$ is increased and the simulation is again started with the initial state as the final state at the previous value of $D$.  At each value of $D$, the simulation is stopped when $|~|\langle \hat{\mathcal{L}}_r \rangle|_{\tau+\Delta\tau} - |\langle \hat{\mathcal{L}}_r \rangle|_{\tau}~|~~ / |\langle \hat{\mathcal{L}}_r \rangle|_{\tau+\Delta\tau} < 10^{-6}$.  The results are shown in Table \ref{chitable}.

\begin{table}[ht]
\caption{Convergence test of $D$ for imaginary time evolution.}
\centering
\begin{tabular}{c c c c}
\hline\hline
D & $\langle n_{\uparrow} \rangle$ & $|\langle\hat{\mathcal{L}}_r\rangle|$ \\ [0.5ex] 
\hline
5& 0.251167 &0.081437 \\
7& 0.251116 &0.081345 \\
9& 0.251163 &0.081198\\
11 & 0.251245 & 0.081079 \\
13 & 0.251319 & 0.081008 \\ 
15& 0.251363 &0.080971 \\
17& 0.251392 &0.080945 \\
19& 0.251410 &0.080928 \\
21& 0.251419 &0.080920 \\
23& 0.251423 &0.080916 \\
25& 0.251425 &0.080913 \\ [1ex]
\hline
\end{tabular}
\label{chitable}
\end{table}

The second test assesses convergence vs. the imaginary timestep size $d\tau$ with a fixed bond dimension of $D=15$.  The test is conducted in the same manner as the first test.  The results are shown in Table \ref{dtautable}. 

\begin{table}[ht]
\caption{Convergence test of $d\tau$ for imaginary time evolution.}
\centering
\begin{tabular}{c c c c}
\hline\hline
$d\tau$ & $\langle n_{\uparrow} \rangle$ & $|\langle\hat{\mathcal{L}}_r\rangle|$ \\ [0.5ex] 
\hline
0.1& 0.251516 &0.080544 \\
0.01& 0.251380 &0.080964 \\
0.001& 0.251367 &0.081004 \\
0.0001 & 0.251365 & 0.081008 \\[1ex]
\hline
\end{tabular}
\label{dtautable}
\end{table}

\subsection{Simulation results for 2-local form of $\hat{\mathcal{L}}_r$}
\begin{figure}
\includegraphics[scale=0.37]{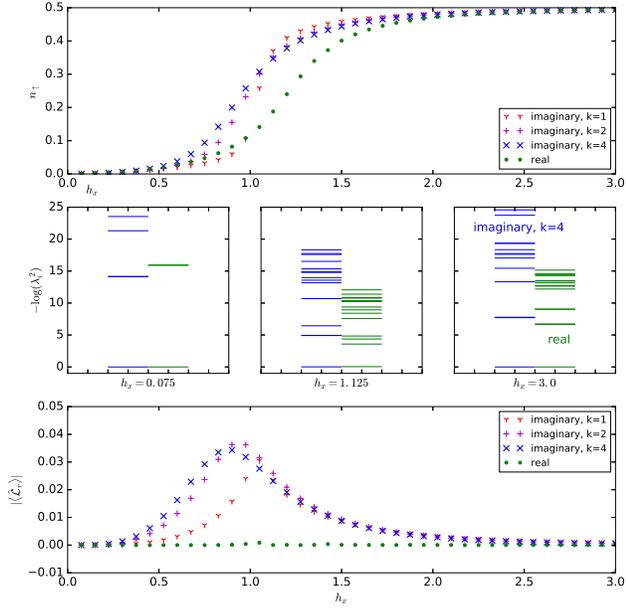}
\caption{(color online).  A comparison of iTEBD imaginary time and iTEBD real time evolution results for the infinite-size dissipative Ising chain defined in the main text. The convergence criterion is given in the text. The dissipation rate is fixed at $\gamma=0.5$.  The imaginary time evolution results are for the 2-local form of $\hat{\mathcal{L}}_r$ given in Eq.(\ref{2local}).  The match between imaginary time evolution and real time evolution is not as good as with the 4-local form of $\hat{\mathcal{L}}_r$ due to fewer intersite couplings in $\mathcal{H}$ with the 2-local form.  (Top) Up-spin density $n_{\uparrow}$ vs. transverse magnetic field strength $h_x$ for real time evolution and different values of $k$ for the imaginary time evolution. Unlike the case in the main text, the overall accuracy of the imaginary time evolution does not improve with larger $k$.  (Middle) Converged entanglement spectra for three different values of $h_x$, with imaginary time evolution results shown for $k=4$. Consistent with an area law, the spectra decay (roughly) exponentially, suggesting proximity to the NESS. (Bottom) Since $\hat{\mathcal{L}}|\rho_{\infty}\rangle=0$, $|\langle \hat{\mathcal{L}}_r\rangle|$ serves as a figure of merit for proximity to the NESS.}
\label{fig_2localupspin}
\end{figure}

For the dissipative Ising model specified in the main text,  we here show a demonstration of the imaginary time method with a 2-local form of $\hat{\mathcal{L}}_r$:
\begin{align}
&\hat{\mathcal{L}}_{r} = -i(H_r\otimes\mathbb{1}-\mathbb{1}\otimes H_r^{\textrm{T}}) \nonumber\\&~~~~~~~+\frac{\gamma}{4}(2\sigma_{-}^{[r]}\otimes\sigma_{-}^{[r]}  - \sigma_{+}^{[r]}\sigma_{-}^{[r]}\otimes\mathbb{1}^{[r]}-\mathbb{1}^{[r]}\otimes \sigma_{+}^{[r]}\sigma_{-}^{[r]}) \nonumber\\&~~~~~~~+\frac{\gamma}{4}(2\sigma_{-}^{[r+1]}\otimes\sigma_{-}^{[r+1]}  - \sigma_{+}^{[r+1]}\sigma_{-}^{[r+1]}\otimes\mathbb{1}^{[r+1]} \nonumber\\ &~~~~~~~~~~~~~~~~-\mathbb{1}^{[r+1]}\otimes \sigma_{+}^{[r+1]}\sigma_{-}^{[r+1]}), \nonumber\\
&H_r=\sigma_z^{[r]}\sigma_z^{[r+1]}+\frac{1}{2}h_x(\sigma_x^{[r]}+\sigma_x^{[r+1]}).
\label{2local}
\end{align}
It is straightforward to verify that this form satisfies $\hat{\mathcal{L}}=\sum_{r\epsilon\mathbb{Z}} \hat{\mathcal{L}}_r$, where $\hat{\mathcal{L}}$ is defined in the main text.  Fig.~\ref{fig_2localupspin} compares converged values of $n_{\uparrow}$, the entanglement spectrum, and $|\langle \hat{\mathcal{L}}_r \rangle|$ from real time evolution to those from imaginary time evolution.  We use the same bond dimension, timestep sizes, and convergence criterion as for the 4-local $\hat{\mathcal{L}}_r$ (see main text). For each value of $h_x$, the same random initial state is used for the real and imaginary time evolutions (for all $k$), which are both performed on a two-site iMPDO. Inside the crossover region, the data shows rough qualitative agreement between the different evolutions.  Comparing with the results in the main text, it is clear that accuracy is greatly enhanced by using the 4-local form of $\hat{\mathcal{L}}_r$ instead of the 2-local form.  This is because the 4-local form results in more inter-site couplings in $\mathcal{H}$, making the groundstate of $\mathcal{H}$ a better approximation for the groundstate $|\rho_{\infty}\rangle$ of $\hat{\mathcal{L}}^{\dagger}\hat{\mathcal{L}}$.

\subsection{Comparison of hybrid method and real time evolution}

In this section we give further details related to Fig. (2) in the main text, where the hybrid method is shown to be efficient where real time evolution becomes inefficient.  The 4-local form of $\hat{\mathcal{L}}_r$ is used (see main text) and the simulation parameters are $h_x=1$, $\gamma=0.01$, $dt=10^{-2}$, and $d\tau=10^{-2}$.  The same hardware (a single CPU laptop) is used for all simulations.

\begin{figure}
\includegraphics[scale=0.46]{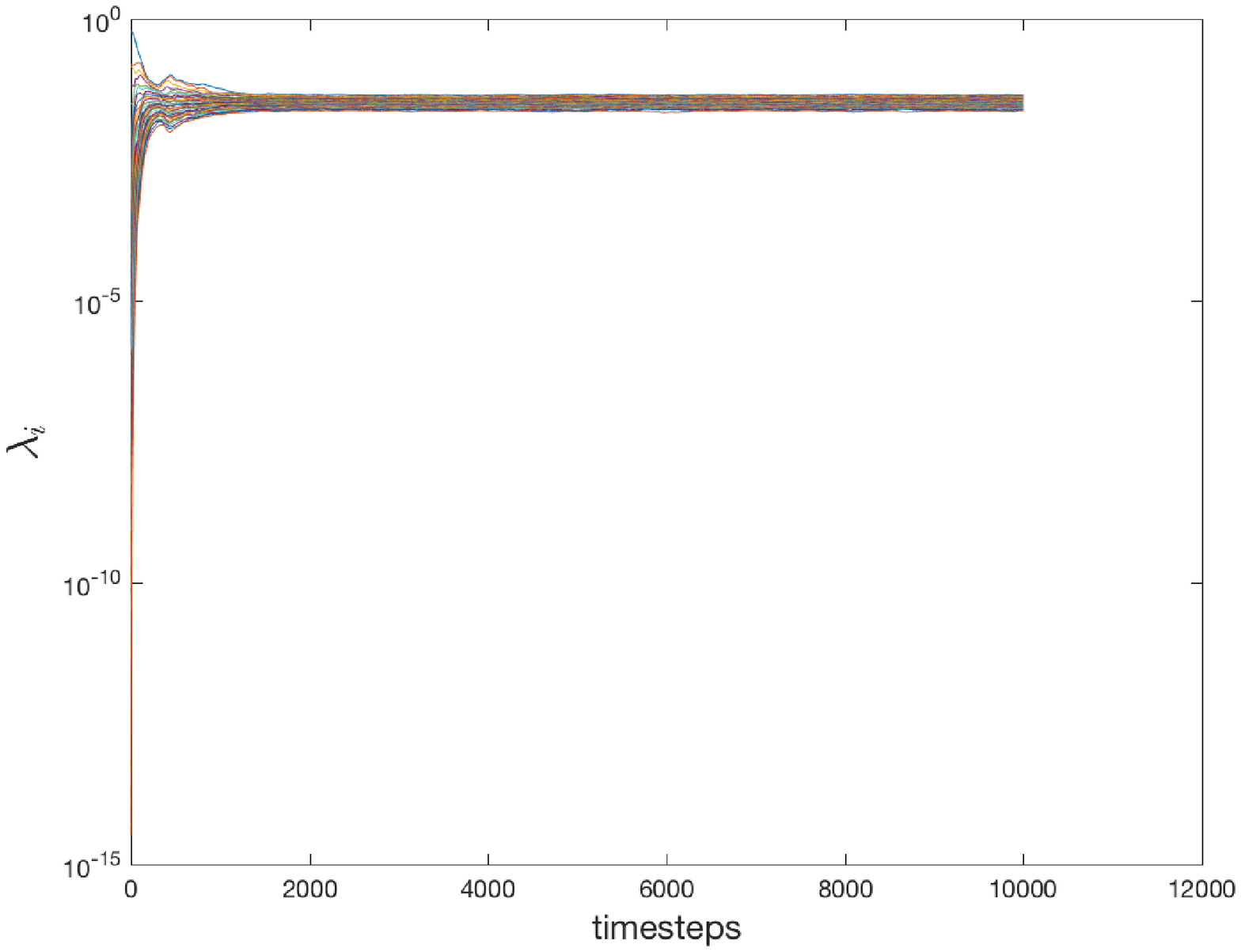}
\caption{(color online) The spectrum ($D=25$) during real time evolution from the initial state rapidly plateaus to a violation of the area law and remains roughly constant over a timescale equal to the largest intrinsic timescale of the system (see text for system and simulation parameters).}
\label{fig_purerealspec}
\end{figure}

For the pure real time evolution simulation, a highly entangled initial state is created with $D=6$ but before the first timestep the bond dimension is increased to $D=25$ with the additional 18 singular values ($\lambda_{i>6}$) initialized at zero.  The simulation is then run for $10^4$ timesteps so that the total physical runtime is the same as the largest timescale ($1/\gamma$) in the system. Fig. (\ref{fig_purerealspec}) shows the time evolution of the spectrum.  It is clear that the system state becomes trapped in an area-law violating portion of the statespace and does not reach the NESS.  The computational time for the simulation is on the order of $10^1$ hours.

\begin{figure}
\includegraphics[scale=0.46]{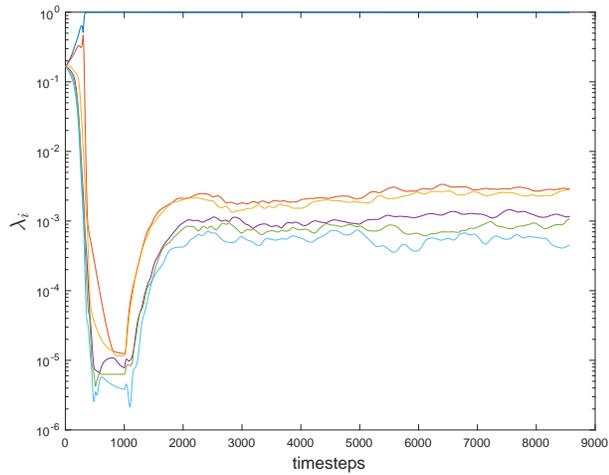}
\caption{(color online) For the hybrid method simulation, the spectrum ($D=6$) rapidly decays to a weakly correlated form.  The imaginary time evolution is over the first 1000 timesteps. See text for system and simulation parameters.}
\label{fig_hybridspec}
\end{figure}

For the hybrid method simulation, the initial state is the same as the one used for the pure real time simulation but the bond dimension is kept at $D=6$ during both imaginary and real time evolution.  The imaginary time evolution is run for $10^3$ timesteps, which takes about one minute, after which real time evolution is performed until $|\langle \hat{\mathcal{L}}_r \rangle|<10^{-6}$, which takes about another eight minutes. The hybrid method thereby achieves $|\langle \hat{\mathcal{L}}_r \rangle|<10^{-6}$ in less than ten minutes of computation time, while pure real time evolution remains stuck at $|\langle \hat{\mathcal{L}}_r \rangle|\approx10^{-1}$ for more than ten hours of computation time.  Fig. (\ref{fig_hybridspec}) shows the evolution of the spectrum during the full hybrid method evolution.

\end{document}